\documentclass[aps,preprint, nofootinbib]{revtex4} 
\usepackage{color}
\usepackage{graphicx}
\begin{document}
\begin{center}
  {\bf\Large  Discrete molecular dynamics simulations } \\
  {\bf\Large  of peptide aggregation } \\
  \bigskip

{ S. Peng$^{1}$, F. Ding$^1$, B. Urbanc$^1$, S. V. Buldyrev$^1$, \\
    L. Cruz$^1$, H. E. Stanley$^1$ and N. V. Dokholyan$^2$} \\
{
  $^1$Center for Polymer Studies and Department of Physics, \\
  Boston University, Boston, MA 02215;  \\
  $^2$ Department of Biochemistry and Biophysics, \\University of North
  Carolina at Chapel Hill,
  Chapel Hill, NC 27599  \\
} \bigskip

\end{center}
\bigskip
\centerline {ABSTRACT}
\bigskip \bigskip 

{ We study the aggregation of peptides using the discrete molecular dynamics
simulations. At temperatures above the $\alpha$-helix melting temperature of
a single peptide, the model peptides aggregate into a multi-layer parallel
$\beta$-sheet structure. This structure has an inter-strand distance of $4.8
\mbox{\AA}$ and an inter-sheet distance of $10 \mbox{\AA}$, which agree with
experimental observations. In this model, the hydrogen bond interactions give
rise to the inter-strand spacing in $\beta$-sheets, while the G${\bar o}$
interactions among side chains make $\beta$-strands parallel to each other
and allow $\beta$-sheets to pack into layers. The aggregates also contain
free edges which may allow for further aggregation of model peptides to form
elongated fibrils.  }

\bigskip
\bigskip
\section{Introduction}
\label{sec: Introduction}

Protein misfolding and polypeptide aggregation are in the focus of
interdisciplinary statistical physics because of their relevance to amyloid
diseases such as Alzheimer's disease, Parkinson's disease and Huntington's
disease. Even though polypeptides related to these diseases share no sequence
or secondary structure similarity, they can aggregate into insoluble fibrils
which share some structural features. These fibrils are typically
$100\mbox{\AA}$ in diameter, and several thousand Angstroms in
length~\cite{Sipe00_JSB}.  X-ray diffraction studies
\cite{Eanes68_JHC,Sunde97_JMB} reveal the common structural features for
these amyloid fibrils: the presence of a 4.7--4.8 $\mbox{\AA}$ inter-strand
spacing along the fibril axis and a 9--10 $\mbox{\AA}$ inter-sheet spacing
perpendicular to the fibril axis~\cite{Bonar69_PSEBM,Kirschner86_PNAS}.
Although advances have been made toward understanding the structural
characteristics of the fibrils and the mechanism of fibril formation, our
knowledge of the detailed fibrillar structure and mechanisms of amyloid
assembly is limited.

Molecular dynamics provides a way to study the aggregation mechanism at the
molecular level. However, continuous all-atom molecular dynamics simulations
with realistic force fields in a physiological solution are not fast enough
to monitor a complete aggregation process from monomers to fully-formed
fibrils.  Recently, a discrete molecular dynamics (DMD)
algorithm~\cite{Zhou97_PNAS,Dokholyan98_FD} using a coarse-grained peptide
model has been successfully implemented to study protein folding
thermodynamics and protein folding kinetics~\cite{Borreguero02_JMB}.  This
computationally fast and dynamically realistic simulation technique has also
been applied to study the aggregation of a small number of Src SH3 domain
proteins~\cite{Ding02_JMB} and the competition of refolding and aggregation
of four-helix bundles~\cite{Smith01_JMB}.

Here we study the aggregation of a large number of peptides. We choose
40-amino acid amyloid $\beta$ peptide
($A\beta_{1-40}$~\cite{Coles98_Biochem}, protein data bank (PDB)~\cite{PDB}
access code 1BA4), which is associated with Alzheimer's disease, to construct
model peptides.  Our results show that model peptides can aggregate into
multi-layer $\beta$-sheet structures with free edges \cite{Richardson02_PNAS}
which may enable further fibrillar elongation.  The computed diffraction
pattern of our simulated multi-layer $\beta$-sheet is consistent with
experimental observations~\cite{Malinchik98_BJ,Serpell00_BBA}.

\section{Methods}
\label{sec: Methods}

\subsection{Two-bead model}
\label{subsec: Model}
\subsubsection{Geometry of model peptide: beads and permanent bonds}

\begin{figure}[hb]
\centering
 \includegraphics*[width=12.0cm]{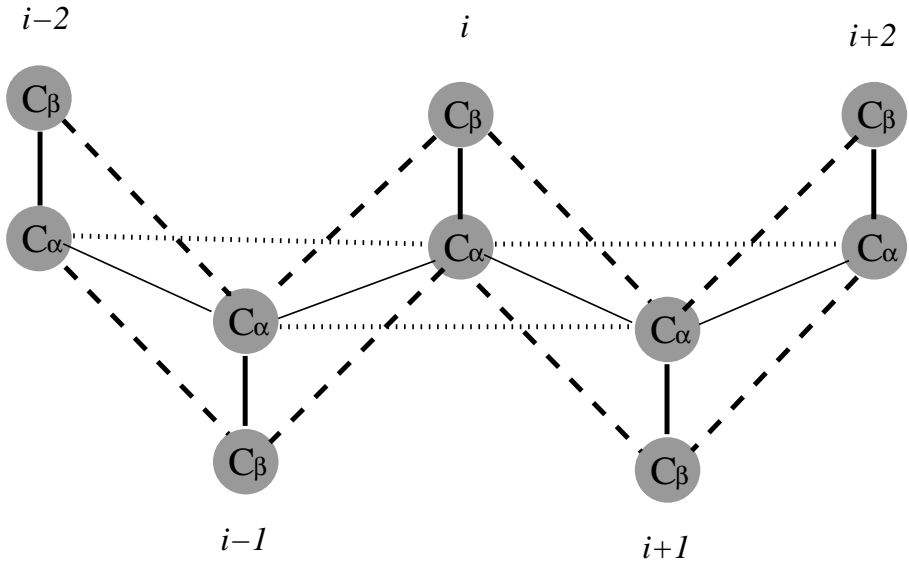} 
\caption{Schematic diagram of two-bead model. Each amino acid in the
  $A\beta_{1-40}$ peptide is represented by two beads: $C_{\alpha}$ bead
  represents backbone atoms and $C_\beta$ bead represents side chain atoms
  ($C_\beta$ is absent for amino acid glycine). The geometry of the peptide
  is modeled by applying permanent bonds among these beads: covalent bonds
  (bold lines), peptide bonds (thin lines) and additional constraints (dashed
  and dotted lines). Interactions between side chains are modeled by G${\bar
    o}$ potentials between $C_\beta$ beads, and interactions between backbone
  atoms are modeled by hydrogen bond interactions between $C_{\alpha}$
  beads.}
\label{fig:Model}
\end{figure}

We model each amino acid in the $A\beta_{1-40}$ peptide by two beads --
$C_\alpha$ bead representing backbone atoms and $C_\beta$ bead representing
side chain atoms (for glycine, $C_\beta$ is absent). Each bead has an index
$i$ indicating the position of amino acid in the sequence starting from the
N-terminus. The geometry of the peptide is modeled by applying permanent
bonds among these beads~\cite{Ding02_BJ}.  These bonds include covalent bonds
between $C_{\alpha i}$ and $C_{\beta i}$, peptide bonds between $C_{\alpha
  i}$ and $C_{\alpha (i \pm 1)}$, additional constraints between $C_{\beta
  i}$ and $C_{\alpha (i\pm1)}$, and also between $C_{\alpha i}$ and
$C_{\alpha (i\pm2)}$ (Fig.~\ref{fig:Model}). These additional constraints are
introduced to model angular constraints between side chains and the backbone.

All permanent bonds are realized by infinitely high potential well interactions
between the related beads~\cite{Dokholyan98_FD}.
\begin{eqnarray}
\label{bonds}
        V^{bond}_{ij} \equiv  \left\{ \begin{array}{ll}
                0 & \mbox{$D_{ij}(1-\sigma_{ij})< |r_i-r_j| <D_{ij}(1+\sigma_{ij})$ };\\
               +\infty & \mbox{otherwise}.\end{array} \right.
\end{eqnarray}
Here $D_{ij}$ is the bond length between beads $i$ and $j$, and $\sigma_{ij}$
is the relative deviation of this bond length. The average lengths for these
bonds can be obtained from statistical analysis of distances within the
$A\beta_{1-40}$ NMR structures~\cite{Coles98_Biochem}.  Table~\ref{Tab:Bonds}
presents the average lengths and their relative deviations~\cite{Deviation}
used in our model.

\subsubsection{Interactions between $C_\beta$ beads: G${\bar o}$ model}
\begin{figure}[ht!]  
  \centering \includegraphics*[width=12.0cm]{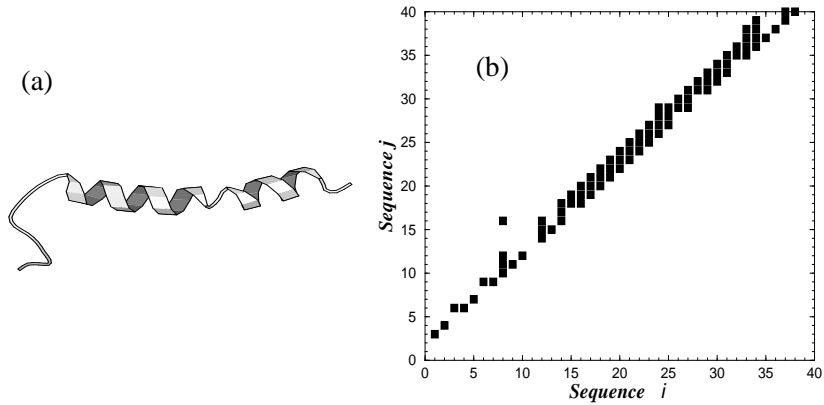}
\caption{ (a) The NMR structure of $A\beta_{1-40}$
peptide~\cite{Coles98_Biochem} used to construct the two-bead model peptide
with G${\bar o}$ potentials and hydrogen bond interactions. The picture is
created with the program {\it Molscript} \cite{Kraulis91_JAC}. (b) The
contact map for structure (a).  Note that the $\alpha$-helical region is from
amino acid 15 to 36 (Q15-V36). }
\label{fig:Native}
\end{figure}

Typically the G${\bar o}$ potentials~\cite{Zhou97_PNAS,Taketomi75_IJPPR} are
used to model proteins with well-defined globular native states. Side chains
which form contacts in the native state (native contacts) experience
attractive G${\bar o}$ potential.  However, $A\beta_{1-40}$ peptide is
``natively unfolded''. NMR studies suggest that in hydrophobic environments
the $A\beta_{1-40}$ peptide assumes mostly $\alpha$-helical
conformation~\cite{Coles98_Biochem}. Fig.~\ref{fig:Native} (a) shows one of
these NMR structures.  Therefore, we apply G${\bar o}$ potentials to preserve
this well-defined, mostly $\alpha$-helical structure of the $A\beta_{1-40}$
peptide.  In our two-bead model a native contact is defined when two
$C_\beta$ beads are closer than $D^{G{\bar o}}=7.5 \mbox{\AA}$ within the NMR
structure of the $A\beta_{1-40}$ peptide.  All the $C_\beta$ beads can not be
closer than the hard-core distance $D_{HC}^{G{\bar o}}=4.5 \mbox{\AA}$.  In
particular, the structure-specific G${\bar o}$ potentials make the side
chains indexed by $i$ within the $\alpha$-helix region of $A\beta_{1-40}$
peptide attract side chains $i\pm2,i\pm3$ and $i\pm4$.  Fig.~\ref{fig:Native}
(b) shows the native contact map for the NMR structure of $A\beta_{1-40}$
peptide shown in Fig.~\ref{fig:Native} (a).

To study the aggregation we need to simulate also the interactions between
different peptides.  We apply G${\bar o}$ potentials for $C_\beta$ beads in
different peptides by an assumption that two amino acids which interact with
each other in a single peptide will interact in the same way in different
peptides.  For example, amino acids 16 and 19 form a native contact in the
NMR structure. Thus, amino acids at 16 and 19 of peptide 1 will experience
attractive G${\bar o}$-type interaction with amino acids 19 and 16 of the
peptide 2, respectively.  The strength of G${\bar o}$ interactions is set to
unity $\epsilon^{G{\bar o}}=1$.

\subsubsection{Interactions between $C_\alpha$ beads: hydrogen bond}

For many globular proteins it has been observed that the number of backbone
hydrogen bonds for each amino acid does not exceed two~\cite{HydrogenBond}.
Also, whenever two hydrogen bonds are formed in a particular peptide block
they are approximately parallel to each other. In order to incorporate these
two facts in our model we introduce two criteria for hydrogen bond formation:
(i) that each $C_\alpha$ bead can form up to two effective hydrogen bonds,
and (ii) that the two hydrogen bonds formed by the same C$\alpha$ bead must
be approximately parallel.

\begin{figure}[ht!]
  \centering \includegraphics*[width=12.0cm]{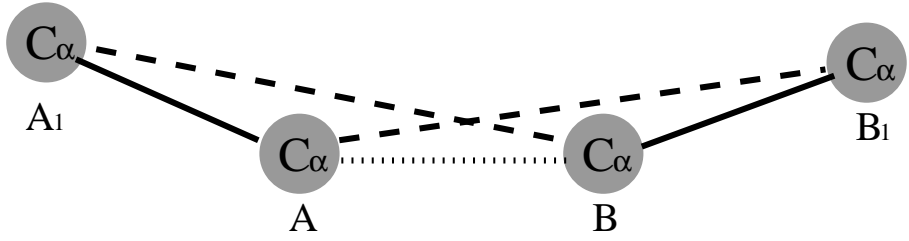}
\caption{Model of a hydrogen bond. Existing hydrogen bonds 
  $AA_1$ and $BB_1$ are shown in bold lines. When the beads $A$ and $B$ come
  to a distance 5$\mbox{\AA}$, a new hydrogen bond (dotted line) may form if
  the distances $A_1B$ and $B_1A$ satisfy inequalities 8.7$\mbox{\AA}$ $\le$
  $A_1B$ $\le$ 10.0$\mbox{\AA}$ and 8.7$\mbox{\AA}$ $\le$ $B_1A$$\le$
  10.0$\mbox{\AA}$. If the bond $AB$ is formed, the auxiliary bonds $A_1B$
  and $B_1A$ (dashed lines) are formed simultaneously.  These bonds can
  fluctuate within the interval 8.7-10$\mbox{\AA}$ and
  cannot be broken unless beads $A$ and $B$ move away from each other to a
  distance 5$\mbox{\AA}$. If the beads $A$ and $B$ have enough kinetic energy
  to leave the hydrogen bond attraction well, their velocities are changed in
  order to conserve energy and momentum, and the hydrogen bond $AB$ is
  destroyed simultaneously with the auxiliary bonds $A_1B$ and $B_1A$.  The
  velocities of $A_1$ and $B_1$ do not change at the moment of forming or
  destroying of hydrogen bond $AB$. Analogously, if one of the hydrogen
  bonds, $A_1A$ or $B_1B$, breaks before hydrogen bond $AB$, the
  corresponding auxiliary bonds $A_1B$ or $B_1A$ also breaks. }
\label{fig:HydrogenBond}
\end{figure}

We set the hydrogen bond interaction range between two $C_\alpha$ beads to
$D^{HB}=5.0$$\mbox{\AA}$, and their hard-core distance to
$D_{HC}^{HB}=4.0$\AA.  We use the following procedure in order to satisfy the
criteria for the hydrogen bond formation: when two $C_\alpha$ beads, $A$ and
$B$, come to a distance $D^{HB}$, we check for any existing hydrogen-bond
partners of $A$ and $B$.  If both beads $A$ and $B$ have no existing hydrogen
partners they can form a hydrogen bond automatically.  If one of the beads,
for example $A$, already has one partner, $A_1$, and the distance between the
bead $A_1$ and the bead $B$ is within the range of 8.7-10$\mbox{\AA}$
(i.e. the angle between vectors $\vec{AA_1}$ and $\vec{AB}$ is within the
range of $120^o$-$180^o$), the bead $A$ can form another hydrogen bond with
bead $B$ provided that either the bead $B$ has no existing hydrogen bonds or
its single hydrogen bond partner, $B_1$, has a distance with bead $A$ in the
range of 8.7-10$\mbox{\AA}$ (see Fig.~\ref{fig:HydrogenBond}).  If one of
beads $A$ and $B$ or both already have two hydrogen bond partners, the pair
will proceed with a hard-core collision without forming a new hydrogen
bond. When a new hydrogen bond is formed between beads $A$ and $B$, new
hydrogen bond partners are recorded for these two beads, and whenever a bead
gets two hydrogen bond partners an auxiliary bond is formed between these two
partners.  Every auxillary bond can fluctuate within the range of
8.7-10$\mbox{\AA}$ to keep two hydrogen bonds within the angle
$120^o$-$180^o$ and it cannot be broken unless one of the two hydrogen bonds
is broken. A hydrogen bond between beads $A$ and $B$ can be broken when these
two beads move away from each other to a distance of $D^{HB}$ and their
kinetic energies are higher than $\epsilon^{HB}$. When a hydrogen bond is
formed or broken, the velocities of the beads $A$ and $B$ change in order to
conserve energy and momentum, such that their kinetic energy increases or
decreases by the value $\epsilon^{HB}$. We set $\epsilon^{HB}=3$ as it was
chosen in Ref~\cite{Ding02_JMB} for Src SH3 domain.

\subsection{Computed diffraction}
\label{subsec: Computed diffraction}
For the typical conformation of $A\beta_{1-40}$ peptide aggregation
structure, we calculate the intensity of diffraction pattern using the
elastic diffraction formula~\cite{Ding02_JMB} in order to compare with
experimental results~\cite{Malinchik98_BJ,Serpell00_BBA}.
\begin{eqnarray}
\label{X-Ray}
I( \vec k_f) = | \sum_{j} exp(i(\vec k_f - \vec k_i) \cdot \vec r_j ) |^2,
\end{eqnarray}
where $\vec k_i$ is the wave vector of the incidental X-ray, $\vec k_f$ is
the wave vector of the outgoing X-ray, $\vec r_j$ is the position vector of
$j$th bead, and the summation is taken over all the $C_\alpha$ and $C_\beta$
beads in the structure.

We chose $x$ axis perpendicular to the $\beta$-sheets, and $y$ axis along the
fibrillar axis which is perpendicular to the $\beta$-strands in the
$\beta$-sheets(Fig.~\ref{fig:28AB_PureGo_XRay} (a)). The incoming X-ray with
1$\mbox{\AA}$ wavelength goes along $z$ axis and the diffraction pattern is
collected on a $x-y$ plane behind the aggregate sample. The deflecting angle,
$\theta = cos^{-1}(\vec k_f \cdot \vec k_i/k^2)$, ranges from 0.05 to 0.25 in
radians in order to detect the periodicity of 4$\mbox{\AA}$ to 20$\mbox{\AA}$
in the aggregate structure.  Since amyloid fibrils consist of bundles of
$\beta$-sheet chains which are twisted along the y-axis, there is no
preferred orientation in the $x-z$ plane in the X-ray diffraction
experiments. We rotate the structure candidate around the $y$ (fibrillar)
axis $n$ times by angle $2 \pi/n$ and add all the diffraction intensities to
obtain a final pattern. We take $n=20$ in the present study.

\section{Results for a single peptide}
\label{sec: Results for single peptide}
As an initial test of our model peptide, we perform DMD simulations of a
single peptide to test whether a peptide with random coil conformation
recovers the observed NMR structure. The model peptide is slowly cooled from
$T_i$=$1.00$, which is high enough to render the peptide as a random coil, to
different target temperatures $T_t$=0.60, 0.55, ..., 0.25. For each target
temperature we make 10 trials starting with different initial conformations.
When $T_t$$\le$0.40, the segment Q15-V36 adopts an $\alpha$-helix or two
pieces of left-handed and right-handed $\alpha$-helices. This artifact is
observed because our simplified two-bead model does not distinguish between
different handedness.  At $T_t$=0.40, the N terminus adopts mostly a random
coil conformation.  As $T_t$$<$0.35, the model peptide starts to approach to
its ground state which is an $\alpha$-helix with a single handedness along
the entire peptide chain.  Therefore, as expected within a certain
temperature range around $T$=0.40 during the cooling process the model
peptide adopts partial $\alpha$-helical conformation similar to the observed
one in NMR experiments.

\begin{figure}[h!]
\begin{center}
$\begin{array}{c@{\hspace{1.5cm}}c}
\includegraphics*[width=7.0cm]{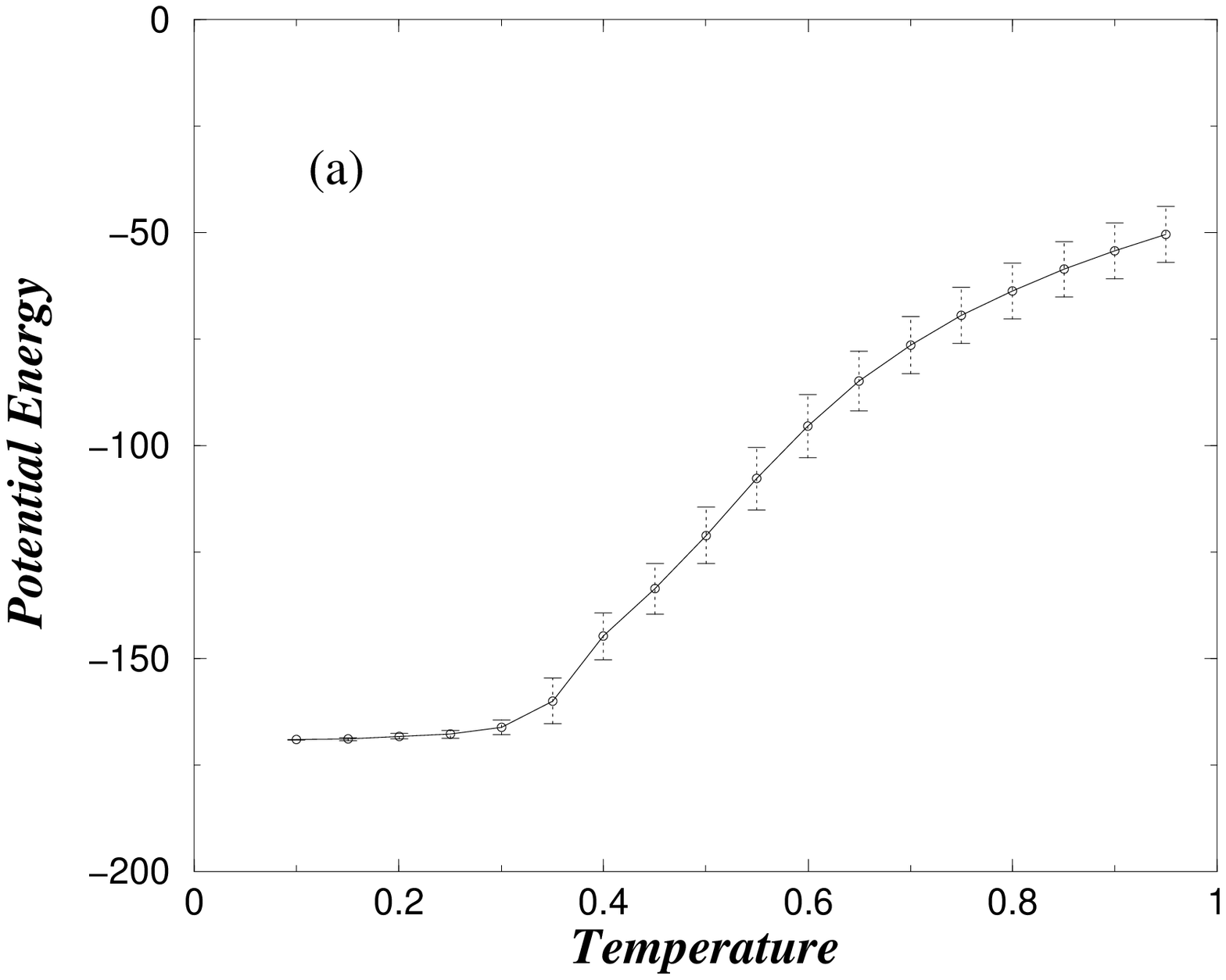} &
\includegraphics*[width=7.0cm]{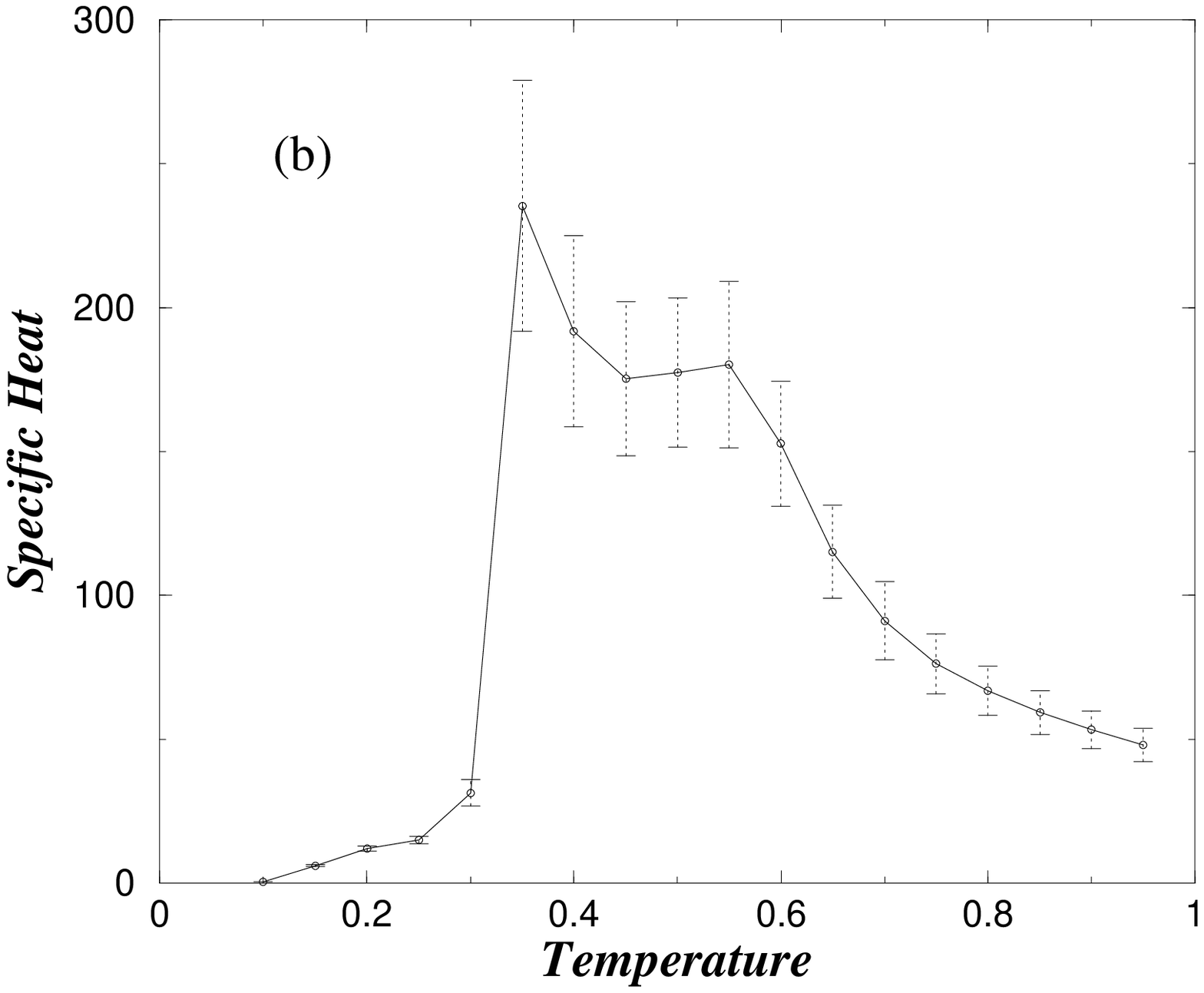} \\
\end{array}$ 
\caption{Temperature dependence of (a) potential energy and 
  (b) specific heat for a single two-bead $A\beta_{1-40}$ model peptide with
  G${\bar o}$ potentials and hydrogen bond interactions. The calculations are
  based on the DMD simulations of $10^7$ time steps for each sampled
  temperature.}
\label{fig:CvT_PureGo}
\end{center}
\end{figure}

We also study the equilibrium behavior of a single model peptide at different
temperatures by measuring the specific heat as a function of temperature.  At
each sampled temperature we start with a ground state conformation and
perform DMD for $10^6$ simulation time steps to equilibrate the system,
followed by additional $10^7$ time steps for the calculations.
Figs.~\ref{fig:CvT_PureGo} (a) and (b) show the potential energy and specific
heat as a function of temperature for a single model peptide, respectively.
The melting of $\alpha$-helix is non-cooperative which can be concluded from
the broad peak between $T_N$$\approx$0.35 and $T_m$$\approx$0.55 in the
specific heat curve (Fig.~\ref{fig:CvT_PureGo} (b)).  $T_N$ corresponds to
the structural transition from an $\alpha$-helix to a random coil for the
first 14 amino acids starting from the N-terminus, while $T_m$ corresponds to
the melting of the $\alpha$-helix in the segment Q15-V36. $T_m$ is higher than
$T_N$ because there are more attractions among the side chains in the segment 
 Q15-V36. 

\section{Results for multiple peptides}
\begin{figure}[h]
\begin{center}
$\begin{array}{c@{\hspace{1.5cm}}c}
\includegraphics*[width=7.0cm]{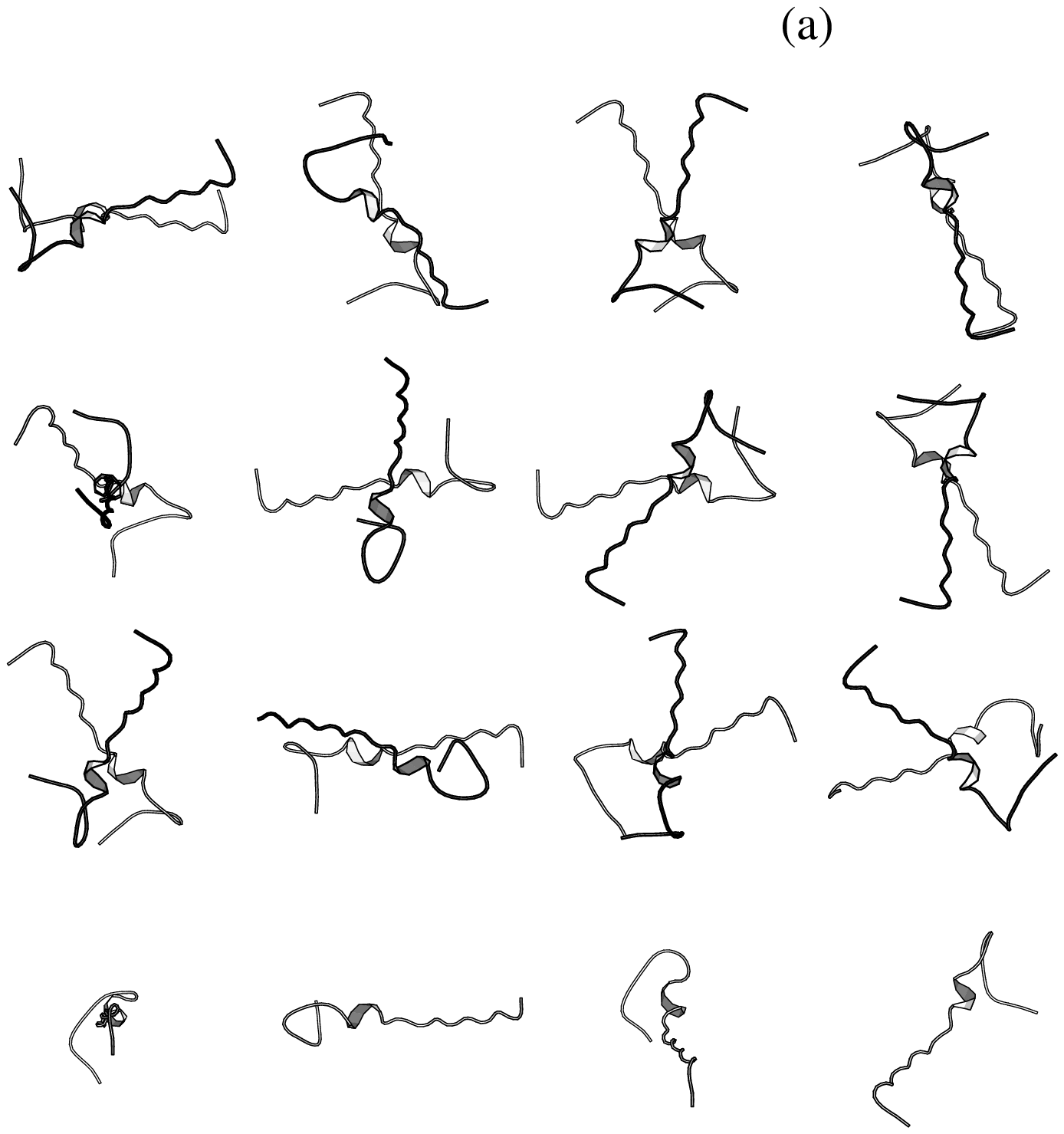} &
\includegraphics*[width=7.0cm]{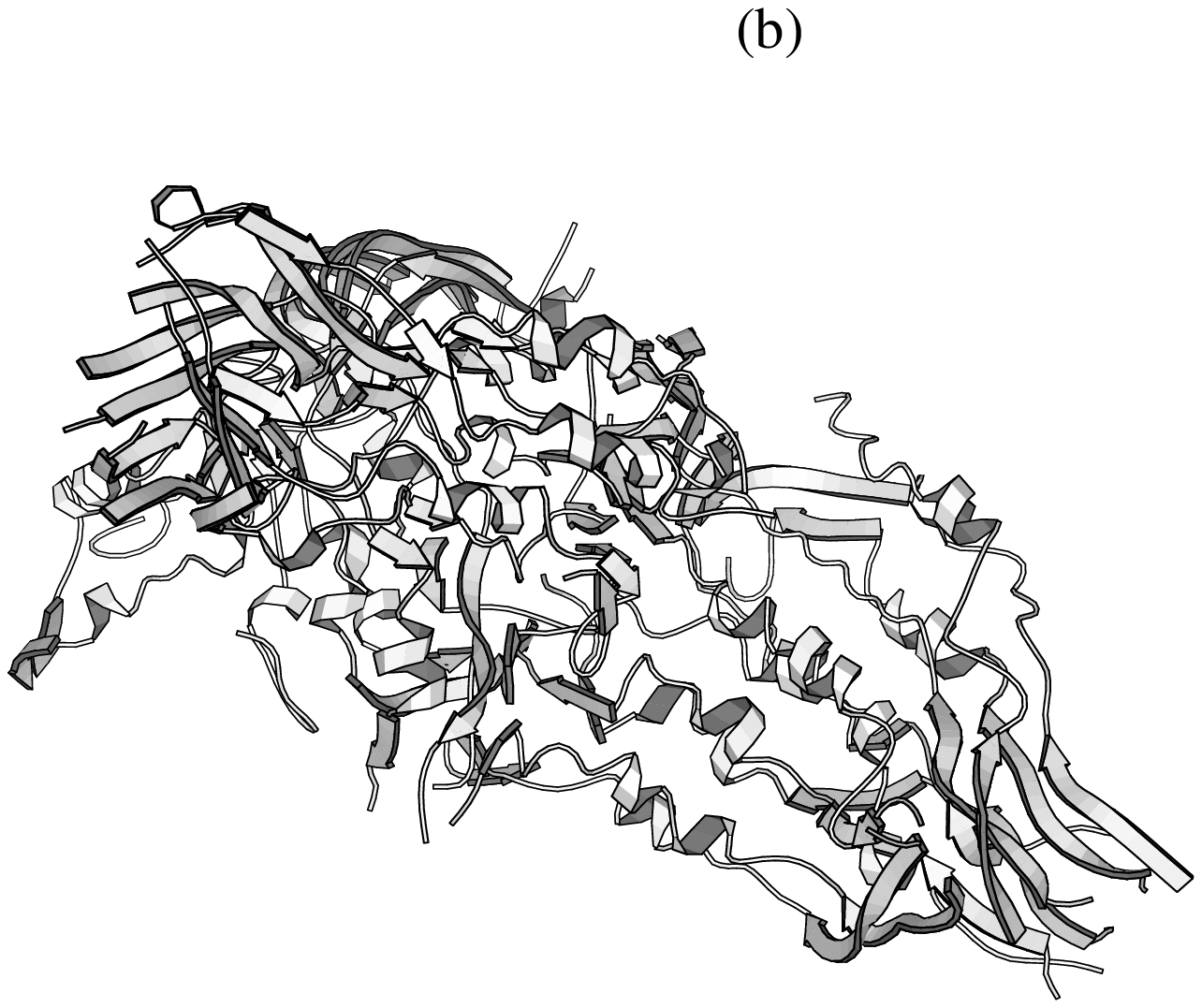} \\
\end{array}$ 
\caption{ DMD simulation of 28 peptides at temperature 0.5: (a) initially,
  all peptides in an original randomly oriented NMR conformation are placed
  on a grid. (b) An amorphous aggregate obtained by DMD simulation at this
  temperature of 0.5. The simulation shows that most of the $\alpha$-helical
  segments are preserved during the aggregation at this temperatures. }
\label{fig:AmorphousStructure}
\end{center}
\end{figure}

In the study of aggregation of many identical peptides, we perform
simulations of 28 peptides in a cubic box with the edge of $200 \mbox{\AA}$
and periodic boundary conditions.  Initially, all the peptides are placed on
a grid and randomly oriented (see Fig.~\ref{fig:AmorphousStructure}
(a)). Then we equilibrate the system at various temperatures: $T_f$=0.4, 0.5,
..., 1.20.

\begin{figure}[h]
\begin{center}
$\begin{array}{c@{\hspace{0.1cm}}c}
\includegraphics*[width=7.0cm]{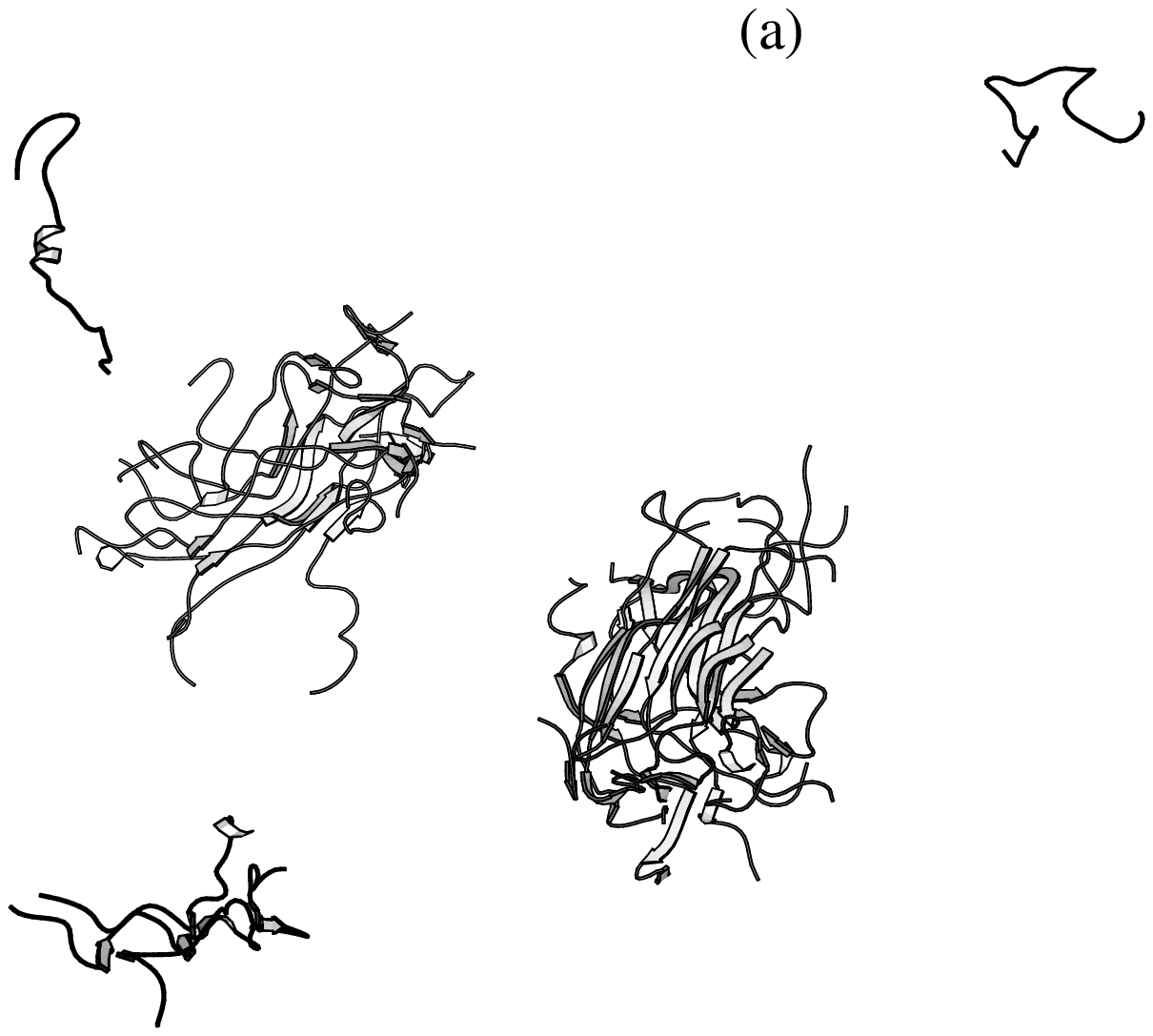} &
\includegraphics*[width=7.0cm]{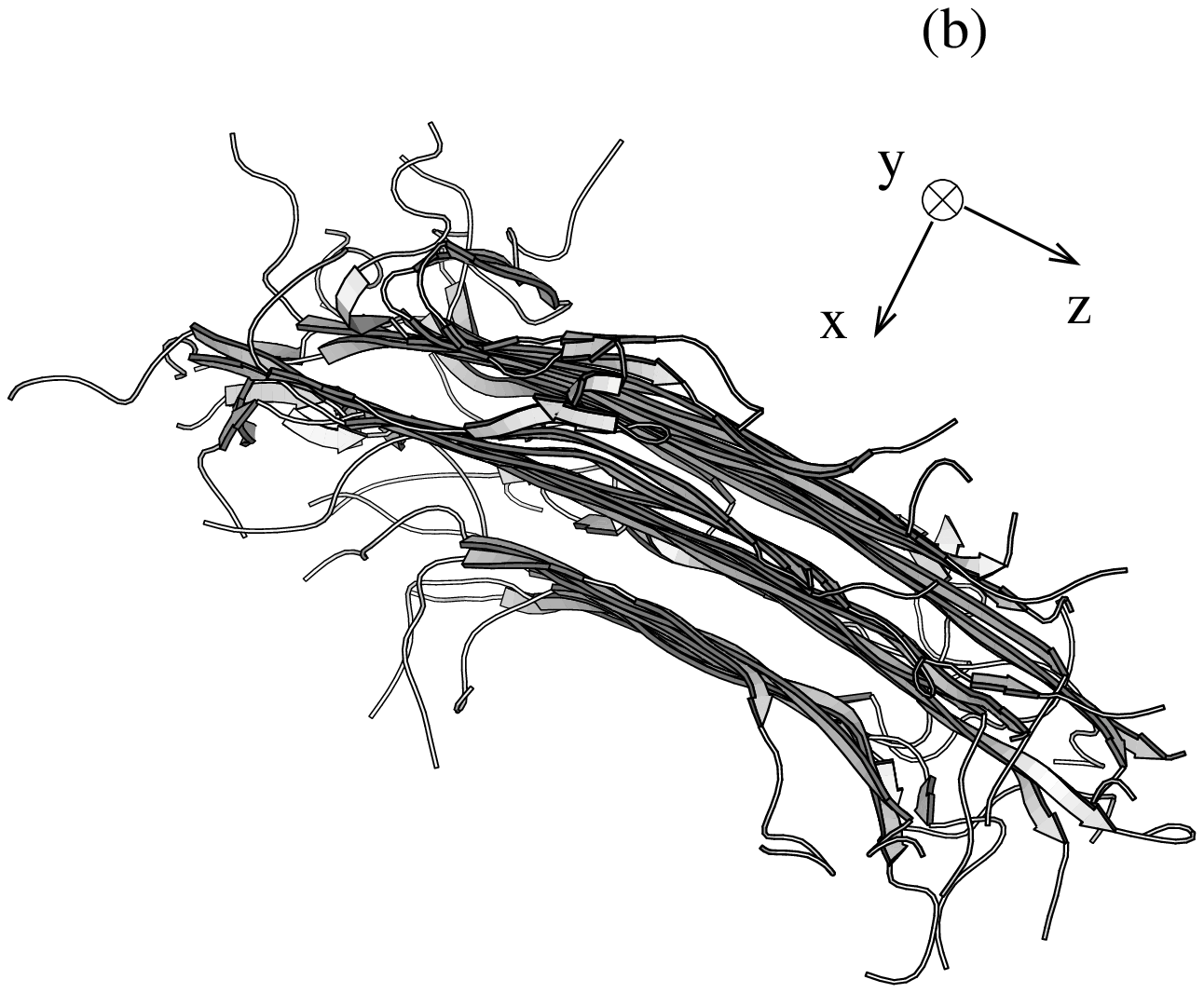} \\
\end{array}$ 
\caption{ DMD simulation of 28 peptides at temperature 0.90. Initial
  conformation is the same as Fig.~\ref{fig:AmorphousStructure} (a). After
  five hundred time steps all peptides acquire random coil conformation
  characteristics for $T$=0.90 (data not shown). (a) Intermediate
  conformation at temperature 0.90 after $10^4$ DMD simulation time steps.
  (b) 3-layer parallel $\beta$-sheet structure formed after $2.5 \times 10^5$
  DMD simulation time steps. This $\beta$-sheet structure contains free edges
  which may allow for further aggregation of model peptides along the y-axis,
  which is perpendicular to the plane of the figure.}
\label{fig:28AB_PureGo}
\end{center}
\end{figure}

\begin{figure}[h]
\begin{center}
$\begin{array}{c@{\hspace{0.5cm}}c@{\hspace{0.5cm}}c}
\includegraphics*[width=5.0cm]{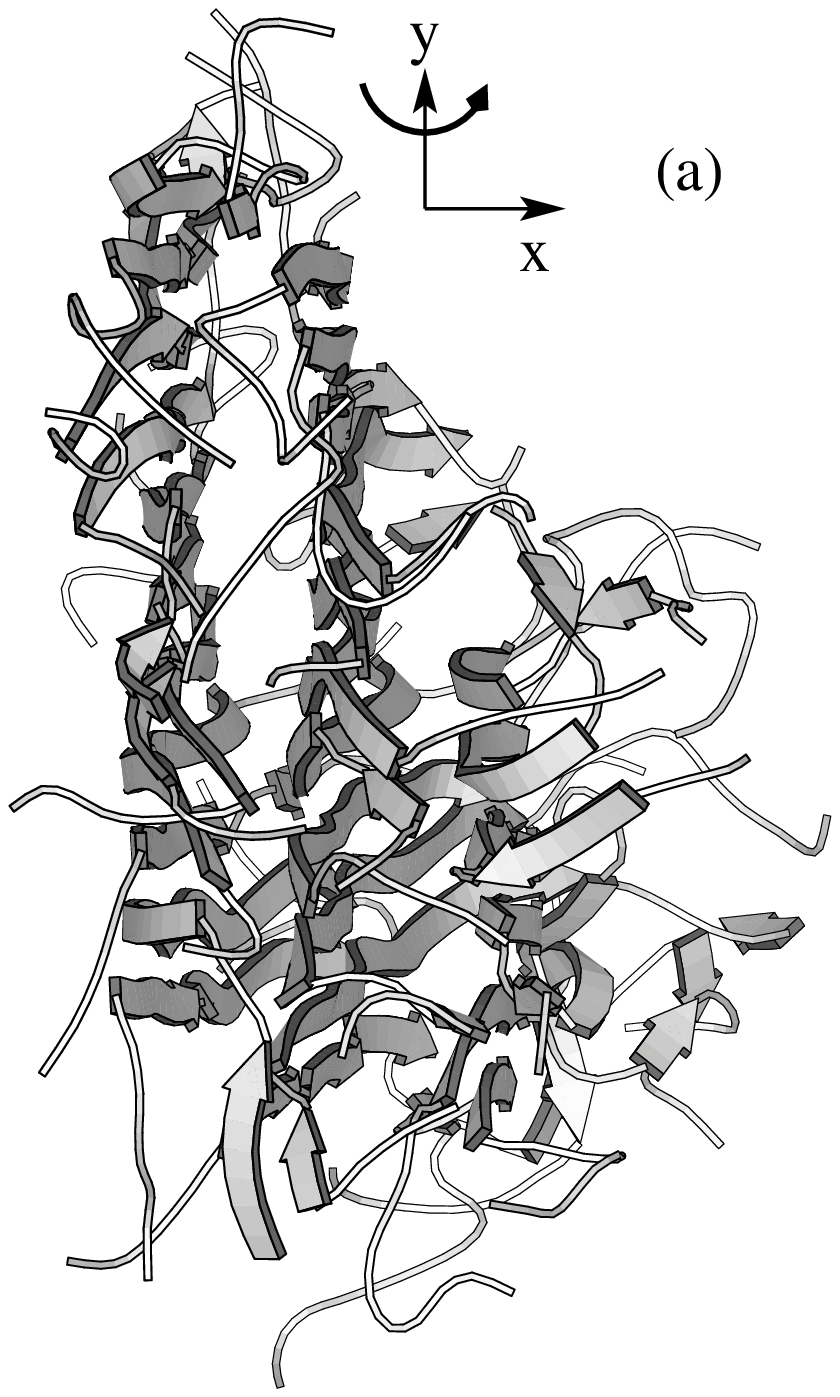} &
\includegraphics*[width=4.0cm]{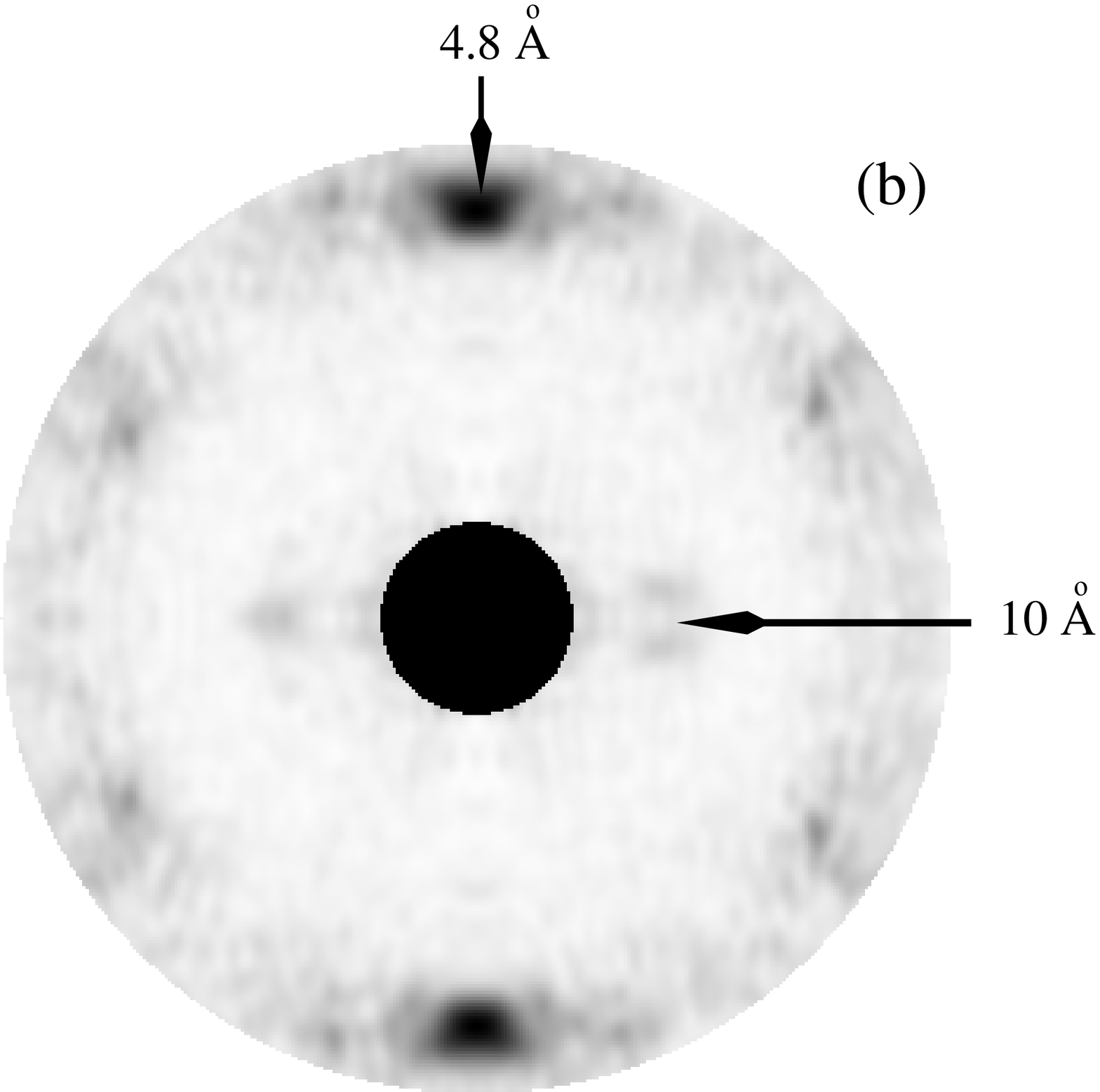} &
\includegraphics*[width=5.0cm]{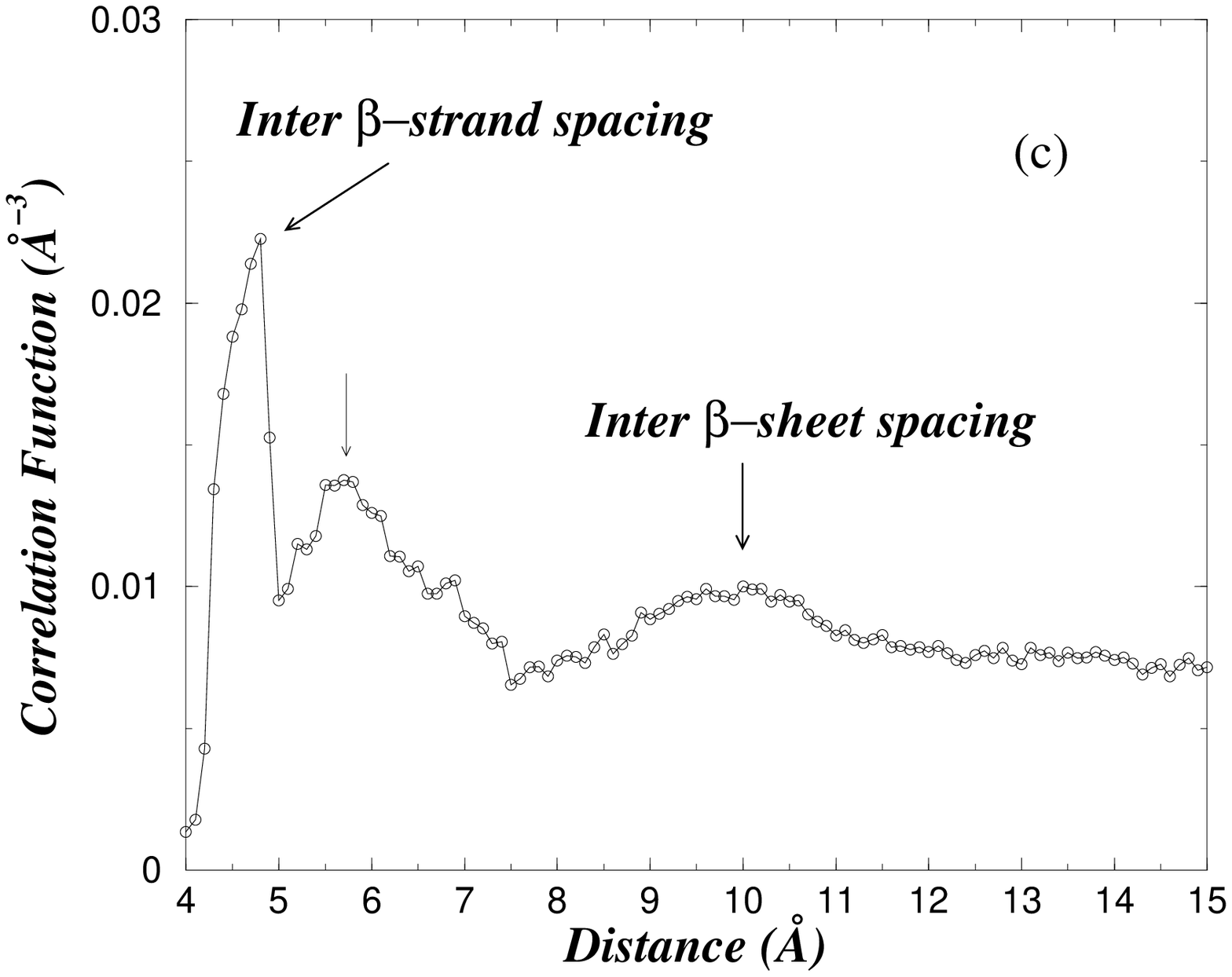} \\
\end{array}$ 
\caption{ (a) The setup of diffraction pattern computation for the 3-layer
  $\beta$-sheet aggregate formed by 28 peptides shown in
  Fig.~\ref{fig:28AB_PureGo}b at a different perspective. The scattering
  occurs along z-axis which is perpendicular to the plane of the figure. (b)
  Computed diffraction pattern collected on a x-y plane behind the
  aggregate. The pattern is averaged over 20 patterns obtained by
  successively rotation of the aggregate around y-axis by $18^o$. (c) Pair
  correlation function for the same aggregate, where peaks around 4.8
  $\mbox{\AA}$ and 10 $\mbox{\AA}$ correspond to inter-strand spacing and
  inter-sheet spacing, respectively.  And the peak around 5.7 $\mbox{\AA}$ is
  mainly from the correlation between neighboring $C_\beta$ beads.
}
\label{fig:28AB_PureGo_XRay}
\end{center}
\end{figure}

At temperatures lower than the melting temperature $T_m$ of a single peptide,
peptides in our model aggregate into amorphous structures where individual
peptides preserve most of their $\alpha$-helical segments as in
Fig.~\ref{fig:AmorphousStructure} (b). When the temperature is higher than
$T_m$, peptides start to aggregate into more ordered structures.  When the
temperature is higher than 1.10, there is no stable aggregate (this threshold
temperature depends on the peptide concentration). At a temperature range
between 0.55 and 1.10, the model peptides can aggregate into multi-layer
$\beta$-sheet structures.  Figs.~\ref{fig:28AB_PureGo} (a) and (b) show the
time evolution of the conformation obtained from DMD simulation at
temperature 0.90.  In Fig.~\ref{fig:28AB_PureGo_XRay} (a) we illustrate the
setup of diffraction computation and in Fig.~\ref{fig:28AB_PureGo_XRay} (b)
we present the calculated diffraction pattern. The relative sharp and intense
4.8 $\mbox{\AA}$ meridional reflection corresponds to the periodic packing of
$\beta$-strands along the fibril axis, and the weaker 10$\mbox{\AA}$
equatorial reflection corresponds to the distance between $\beta$-sheets. In
Fig.~\ref{fig:28AB_PureGo_XRay} (c) we show the calculated pair correlation
function for the same $\beta$-sheet structure.  The peaks around $4.8
\mbox{\AA}$ and 10$\mbox{\AA}$ correspond to the average inter-strand and
inter-sheet spacings, respectively.

\begin{figure}[h]
\begin{center}
$\begin{array}{c@{\hspace{1.5cm}}c}
\includegraphics*[width=7.0cm]{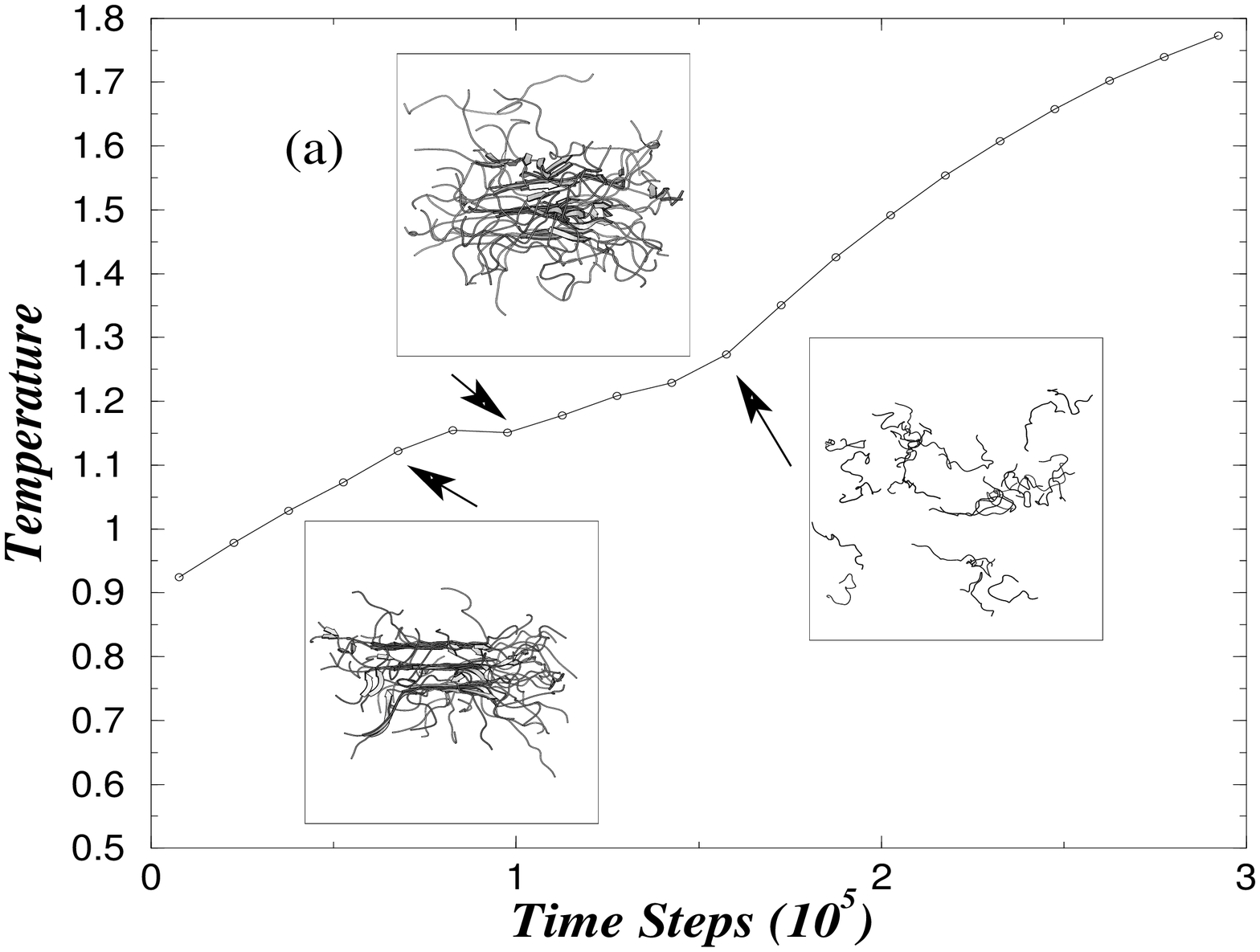} &
\includegraphics*[width=7.0cm]{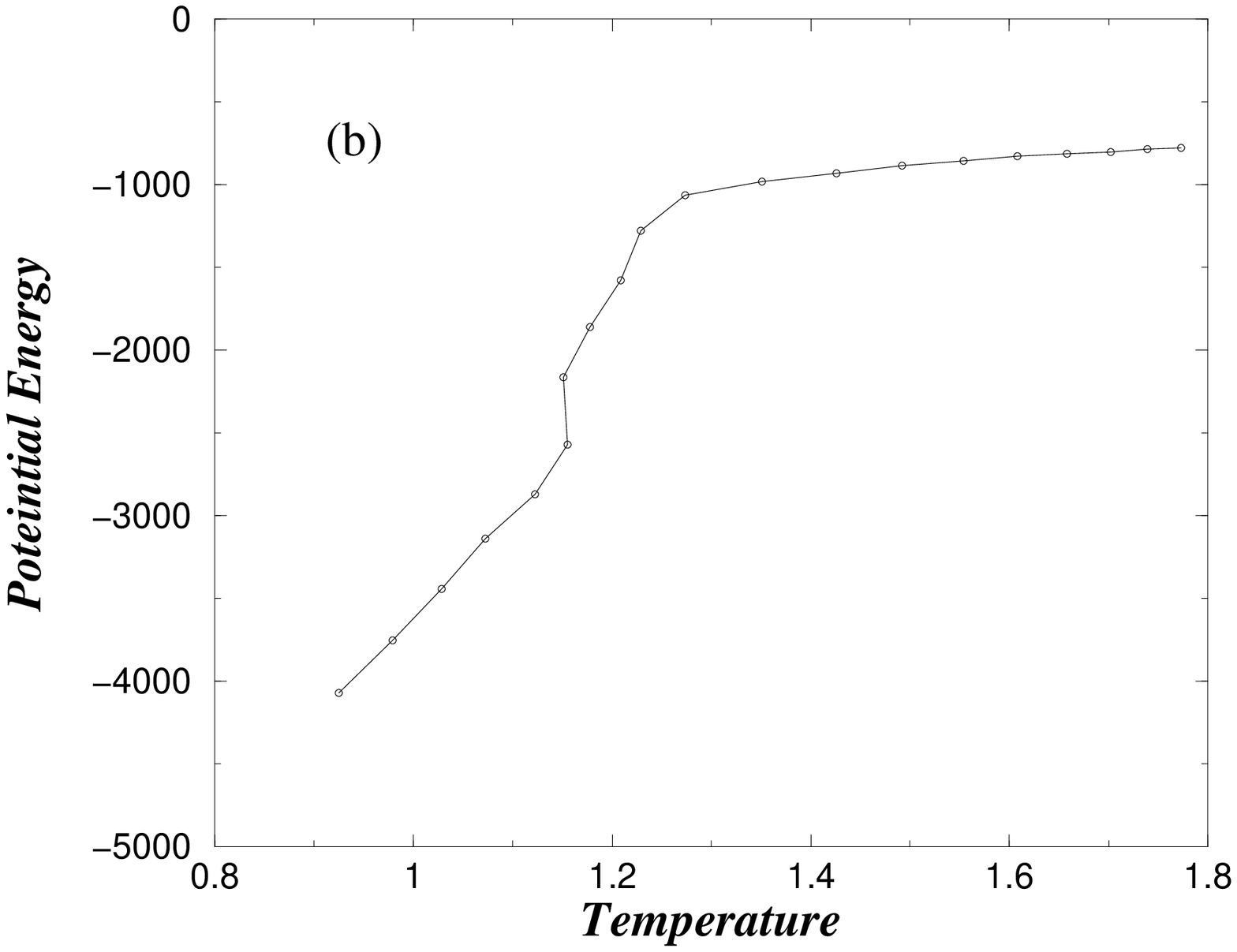} \\
\end{array}$ 
\caption{The melting of the 3-layer $\beta$-sheet structure of 28 $A\beta$
  peptides (Fig.~\ref{fig:28AB_PureGo} (c)).  (a) Time evolution of the
  temperature when the 3-layer $\beta$-sheet is warmed up slowly from
  temperature 0.90 to 2.00. The dissociation temperature is $T_d$=$1.20 \pm
  0.05$. The insets are the conformations at different times/temperatures.
  Note that the third one shows a completely dissociated conformation in 3-d.
  (b) The temperature dependence of potential energy.}
\label{fig:28AB-Melting}
\end{center}
\end{figure}

In order to study the thermostability of this 3-layer $\beta$-sheet
structure, we slowly increase the temperature to $T$=2.0 which is higher
enough to melt the aggregate.  Figs.~\ref{fig:28AB-Melting} (a) and (b) show
the time evolution of temperature and the temperature dependence of potential
energy of the system during melting and dissociation of the $\beta$-sheet
structure, respectively.  As temperature increases from 0.90, the aggregate
becomes less stable.  At temperature around $T$=$1.15\pm0.05$, aggregate
starts to dissociate.  At temperatures higher than $T_d$=$1.20\pm0.05$ the
dissociation completes.

If we assume that the temperature 0.9 at which the aggregation of
$\beta$-sheet is observed corresponds to physiological temperature, 310 K.
At this temperature our single model peptide exists in a random coil
conformation, which corresponds to experimental
observations~\cite{Gursky99_BBA,Soto95_JBC} that in aqueous solution at
physiological temperatures $A\beta_{1-40}$ peptides adopt mostly
$\beta$-sheet and coil conformations.  The temperature of the $\beta$-sheet
dissociation, 1.2, corresponds to 413 K. Temperature $T$=0.40 at which our
model peptide acquires $\alpha$-helical conformation corresponds to very low
physical temperatures which can not be observed experimentally.

\section{Discussion and Conclusion}
\label{sec: Discussion}

In the test of our coarse-grained model of $A\beta_{1-40}$ peptide, we find
that the model peptide most resembles the NMR structure of $A\beta_{1-40}$
peptide around $T$=0.40.  The existence of an optimal temperature range for
protein refolding is also observed in experiments~\cite{Jeanick97_APC} and
other coarse-grained models~\cite{Smith01_PSFG}.  Below $T$=0.40 the
N-terminal region of our model peptide mostly adopts an $\alpha$-helical
conformation.  However, in the present study of aggregation we are focused on
temperatures above 0.40 as the peptides are generally partially or
completely unfolded to initiate the aggregation
process~\cite{Chiti03_Nature}.

In studies of multiple peptides, we demonstrate that peptides aggregate into
amorphous structures (Fig.~\ref{fig:AmorphousStructure} (b)) around $T$=0.50
or multiple-layer $\beta$-sheet structures (Fig.~\ref{fig:28AB_PureGo} (b))
around $T$=0.90. In the amorphous structures, individual peptides tend to
preserve most of the $\alpha$-helical structure, while in the $\beta$-sheet
structures the $\beta$-strands tend to be parallel.  Since the G${\bar o}$
interaction for an $\alpha$-helix favors the formation of contacts between
amino acids $i$ and $i\pm3$, the aggregates with a parallel alignment have
lower potential energies. 

There are caveats due to the simplicity of the two-bead model used in our
study. Each amino acid is represented by only two beads, which do not allow
for an accurate description of the backbone. The backbone in this model is
too flexible, which introduces some artifacts into conformations of
aggregates composed of small number of peptides at low temperatures, such as
dimers, trimers and tetramers.
  
An additional problem is that the chiral symmetry of each amino acid is not
considered in this model. As a result, we observe two $\alpha$-helices with
opposite handedness. As the $A\beta_{1-40}$ NMR structure consists of two
$\alpha$-helices and a hinge in between, there are four low energy states
with combinations of different handedness within the region of
$\alpha$-helices at $T$=0.40. The conformations with mixed handedness appear
with lower probabilities since they have higher potential energies due to the
loss of native contacts and hydrogen bonds in between the two
$\alpha$-helices of different handedness.

Also, due to the simplicity of the two-bead model, we do not account for
specific structural features of $A\beta_{1-40}$ peptides, such as the salt
bridge between D23 and K28~\cite{Petkova02_PNAS}. For the same reason, we can
not expect to explain the differences in aggregation pathways between
$A\beta_{1-40}$ and $A\beta_{1-42}$ alloforms~\cite{Bitan03_PNAS}, nor study
subtle aggregation differences due to amino acid
substitutions~\cite{Bitan03_JBC}.  We show that the DMD algorithm using a
simplified peptide model can reproduce the formation of $\beta$-sheet
structures of 28 peptides with free edges for further fibrillization.
Our study shows that it is possible to investigate in detail the aggregation
of several dozens of peptides using DMD simulations and the coarse-grained
model for peptide structure. Both the number of peptides and the complexity
of the model~\cite{Ding03_PSFG} can be significantly increased within
realistic computational constraints. Thus we regard this study as a first
step toward developing a realistic model of $A\beta$ peptide aggregation.

\section{ACKNOWLEDGMENTS}
\label{sec: ACKNOWLEDGMENTS}
We thank Dr. C.K. Hall, S. Yun, J.M. Borreguero and A. Lam for discussions,
and the Memory Ride Foundation for support. N.V.D. acknowledges the support
of the UNC-CH Research Council Grant.


\newpage
\begin{table}[ht]
\caption{Permanet bonds in the two-bead model}
\label{Tab:Bonds}
\begin{tabular}{l c c}
\toprule
Bond         & Bond length ($\mbox{\AA}$) & ~~Deviation (\%) \\
\colrule
$C_{\alpha i}$ -- $C_{\beta i}$           & 1.55 & 2.4 \\
$C_{\alpha i}$ -- $C_{\alpha (i \pm 1)}$  & 3.82 & 3.1 \\ 
$C_{\beta i}$  -- $C_{\alpha (i\pm1)}$    & 4.66 & 6.5 \\
$C_{\alpha i}$ -- $C_{\alpha (i\pm2)}$    & 5.65 &  14.8 \\
\botrule
\end{tabular}
\end{table}

\begin{thebibliography}{99}

\bibitem{Sipe00_JSB}
J.D. Sipe and A.S. Cohen, 
J. Struct. Biol. {\bf 130,} 88 (2000).

\bibitem{Eanes68_JHC}
E.D. Eanes and G.G. Glenner, 
J. Histochem. Cytochem. {\bf 16,} 673 (1968).

\bibitem{Sunde97_JMB}
M. Sunde, L.C. Serpell, M. Bartlam, P.E. Fraser, M.B. Pepys and C.C. Blake, 
J. Mol. Biol. {\bf 273,} 729 (1997).

\bibitem{Bonar69_PSEBM}  
L. Bonar, A.S. Cohen and M.M. Skinner,
Proc. Soc. Exp. Biol. Med. {\bf 131,} 1373 (1969).

\bibitem{Kirschner86_PNAS} 
D.A. Kirschner, C. Abraham and D.J. Selkoe,
Proc. Natl. Acad. Sci. U.S.A. {\bf 83,} 503 (1986).

\bibitem{Zhou97_PNAS} 
Y. Zhou and M. Karplus, 
Proc. Natl. Acad. Sci. U.S.A. {\bf 94,} 14429 (1997).

\bibitem{Dokholyan98_FD}
N.V. Dokholyan, S.V. Buldyrev, H.E. Stanley and E.I. Shakhnovich,
{Folding $\&$ Design} {\bf 3,} 577 (1998).

\bibitem{Borreguero02_JMB} 
J.M. Borreguero, N.V. Dokholyan, S.V. Buldyrev, E.I. Shakhnovich and H.E. Stanley,
{J.  Mol. Biol.} {\bf 318,} 863 (2002).

\bibitem{Ding02_JMB}
F. Ding, N.V. Dokholyan, S.V. Buldyrev, H.E. Stanley and E.I. Shakhnovich,
{J.  Mol. Biol.} {\bf 324,} 851 (2002).

\bibitem{Smith01_JMB}
A.V. Smith and C.K. Hall,
{J.  Mol. Biol.} {\bf 312,} 187 (2001). 

\bibitem{Coles98_Biochem} 
  M. Coles, W. Bicknell, A.A. Watson, D.P. Fairlie and D.J.Craik,
Biochem. {\bf 37,} 11064 (1998).

\bibitem{PDB}   
  H.M. Berman, J. Westbrook, Z. Feng, G. Gilliland, T.N. Bhat, H. Weissig,
  I.N. Shindyalov and P.E. Bourne,
  Nucleic Acids Research {\bf 28,} 235 (2000).
  
\bibitem{Richardson02_PNAS}
J.S. Richardson and D.C. Richardson,
Proc. Natl. Acad. Sci. U.S.A. {\bf 99,} 2754 (2002).

\bibitem{Malinchik98_BJ}
S.B. Malinchik, H. Inouye, K.E. Szumowski and D.A. Kirschner,
Biophys. J. {\bf 74,} 537 (1998).

\bibitem{Serpell00_BBA}  
L.C. Serpell, 
 {Biochim. Biophys. Acta.} {\bf 1502,} 16 (2000).

\bibitem{Ding02_BJ}
F. Ding, N.V. Dokholyan, S.V. Buldyrev, H.E. Stanley and E.I. Shakhnovich, 
{Biophys. J.} {\bf 83,} 3525 (2002).

\bibitem{Deviation} 
  In a DMD simulation, the speed of the simulation depends on the number of
  collisions.  If a bond width is too small, most of the computing time will
  be wasted on the collisions due to small local vibrations of this bond.
  However, if the bond width too big, the model would not be realistic
  enough.

\bibitem{Taketomi75_IJPPR}
H. Taketomi, Y. Ueda and N. G${\bar o}$, 
Int. J. Pept. Protein Res. {\bf 7,} 445 (1975).

\bibitem{HydrogenBond}  
  Bifurcated hydrogen bonding is very rare and is not considered here.

\bibitem{Gursky99_BBA} 
  O. Gursky and S. Aleshkov,
Biochimica et Biophysica Acta, {\bf 1476,} 93 (2000).

\bibitem{Soto95_JBC}  
  C. Soto, E.M. Castano, B. Frangione and N.C. Inestrosa,
  J. Biol. Chem. {\bf 270,} 3063 (1995).

\bibitem{Jeanick97_APC}
R. Jaenicke and R. Seckler,  
Adv. Protein. Chem.  {\bf 50,} 1 (1997).

\bibitem{Smith01_PSFG} 
A.V. Smith and C.K. Hall,
Proteins: Structure, Function, and Genetics, {\bf 44,} 376 (2001).

\bibitem{Chiti03_Nature} 
F. Chiti, M. Stefani, N. Taddei, G. Ramponi and C.M. Dobson,
Nature {\bf 424,} 805 (2003).

\bibitem{Petkova02_PNAS}
A.T. Petkova, Y. Ishii, J.J. Balbach, O.N. Antzutkin, R.D. Leapman,
F. Delaglio and R. Tycko,    
Proc. Natl. Acad. Sci. U.S.A. {\bf 99,} 16742 (2002).

\bibitem{Bitan03_PNAS}
G. Bitan, M.D. Kirkitadze, A. Lomakin, S.S. Vollers, G.B. Benedek and
D.B. Teplow,  
Proc. Natl. Acad. Sci. U.S.A. {\bf 100,} 330 (2003).

\bibitem{Bitan03_JBC}
G. Bitan, S.S. Vollers and D.B. Teplow,
J. Biol. Chem. {\bf 278,} 34882 (2003).

\bibitem{Ding03_PSFG} 
F. Ding, J.M. Borreguero, S.V. Buldyrev, H.E. Stanley and N.V. Dokholyan,
Proteins: Structure, Function, and Genetics, {\bf 53,} 220 (2003).

\bibitem{Kraulis91_JAC}
P.J. Kraulis, 
J. Appl. Crystallog. {\bf 24,} 946 (1991).

\end{thebibliography}
\end{document}